\documentclass[aps,prl,twocolumn,superscriptaddress,preprintnumbers,10pt]{revtex4-2}
\usepackage{style}

\begin{document}

\title{First-principles upper bounds on dark matter-electron scattering rates from condensed matter sum rules}

\author{Bradford A. Barker}
\email{bbarker@floridapoly.edu}
\affiliation{Department of Physics, Florida Polytechnic University, Lakeland, FL 33805, USA}
\author{Jay Epstein}
\email{jaydepstein@gmail.com}
\author{Luke James}
\email{luke.james@mail.utoronto.ca}
\author{Yonatan Kahn\,\orcidlink{0000-0002-9379-1838}}
\email{yf.kahn@utoronto.ca}
\affiliation{Department of Physics, University of Toronto, Toronto, ON M5S 1A7, Canada}
\author{Elizabeth A. Peterson}
\email{epeterson@lanl.gov}
\affiliation{Theoretical Division, Los Alamos National Laboratory, Los Alamos, NM 87545 USA}
\author{Anirudh Prabhu\,\orcidlink{0000-0001-9115-7844}}
\email{aprabhu@berkeley.edu}
\affiliation{Department of Physics, Princeton University, Princeton, NJ 08544, USA}
\affiliation{Leinweber Institute for Theoretical Physics, University of California, Berkeley, CA 94720, USA}
\affiliation{Theoretical Physics Group, Lawrence Berkeley National Laboratory, Berkeley, CA 94720, USA}
\author{Tanner Trickle\,\orcidlink{0000-0003-1371-4988}}
\email{ttrickle@illinois.edu}
\affiliation{Department of Physics, Grainger College of Engineering, University of Illinois Urbana-Champaign, Urbana, IL 61801, USA}
\author{Samuel L. Watkins\,\orcidlink{0000-0003-0649-1923}}
\email{samuel.watkins@pnnl.gov}
\affiliation{Pacific Northwest National Laboratory, Richland, WA 99354, USA}
\affiliation{Physics Division, Los Alamos National Laboratory, Los Alamos, NM 87545 USA}

\begin{abstract}
A wide variety of condensed matter systems are used or proposed as detectors to search for dark matter-electron scattering. In general, the scattering rate depends on detailed knowledge of the electronic properties of these systems. However, when dark matter couples to electron density, the dark matter-electron scattering rate can be related to the electron energy loss function, whose integrals are bounded by first-principles sum rules that rely on only a few macroscopic target properties.
In this paper, we use these first-principles sum rules to derive upper bounds on the dark matter-electron scattering rate depending on only a few material properties: the plasma frequency $\omega_\text{p}$, the target mass density $\rho_T$, and the static (longitudinal) dielectric function at finite momentum transfer, $\varepsilon(q, 0)$. The bulk material properties $\omega_\text{p}$ and $\rho_T$ vary only over a limited range across a wide variety of materials, 
and to a good approximation, the generic large-$q$ dependence of $\varepsilon(q, 0)$ can be understood from a simple scaling law depending only on $\omega_\text{p}$ which we verify with analytic and numerical examples.
Thus, our upper bounds are largely material-agnostic, and place a fundamental limit on the sensitivity of any dark matter-electron direct detection experiment probing the coupling to electron density.
\end{abstract}

\maketitle

As large-scale WIMP experiments approach the neutrino fog in their search for dark matter (DM)-induced nuclear recoils~\cite{Akerib:2022ort}, enormous progress is being made in new experiments sensitive to DM-electron scattering, especially for DM lighter than the proton~\cite{Essig:2022dfa}. 
While DAMIC-M, using a Si-based detector, is currently the most sensitive to DM-electron scattering in the MeV--GeV range~\cite{DAMIC-M:2025luv}, many other experiments using semiconductors~\cite{CDEX:2022kcd,EDELWEISS:2020fxc,SENSEI:2023zdf,SuperCDMS:2020ymb,SuperCDMS:2025dha}, conventional superconductors \cite{QROCODILE:2024nqm,Hochberg:2026mdl}, and noble liquids~\cite{DarkSide:2022knj,PandaX:2022xqx,XENON:2026qow} are leveraging their mature fabrication and purification pipelines to aid in the search for DM.
Additionally, there are a plethora of proposals for novel detector materials, including polar materials~\cite{Knapen:2017ekk,Griffin:2019mvc,TESSERACT:2025tfw},  graphene~\cite{Hochberg:2016ntt,Catena:2023qkj,Catena:2023awl,Das:2023cbv,Sherpa:2026tgy}, carbon nanotubes~\cite{Cavoto:2017otc,Cavoto:2017otc}, narrow-gap semiconductors~\cite{Hochberg:2017wce,Coskuner:2019odd,Geilhufe:2019ndy,Inzani:2020szg,Chen:2022pyd,Abbamonte:2025guf,Griffin:2025wew}, scintillators~\cite{Derenzo:2016fse,Blanco:2019lrf,Blanco:2021hlm}, quantum dots~\cite{Blanco:2022cel}, and doped semiconductors~\cite{Du:2022dxf} (see Refs.~\cite{Kahn:2021ttr,Essig:2022dfa,Zurek:2024qfm} for reviews of approaches to sub-GeV DM detection).
Thus far, the search for new detectors has largely proceeded by identifying an interesting candidate material based on properties relevant for DM scattering, computing its response~\cite{Essig:2011nj,Essig:2015cda,Griffin:2019mvc,Griffin:2021znd,Hochberg:2021pkt,Knapen:2021run,Boyd:2022tcn,Trickle:2022fwt,Dreyer:2023ovn,Catena:2024rym,Krnjaic:2024bdd,Hochberg:2025rjs,Dreyer:2026bmz}, and estimating a sensitivity.

In this \textit{Letter} we take an alternative approach and address the following question: given a specific DM-electron interaction, what is the \emph{largest possible} DM-electron scattering rate? A material which saturates this upper bound would be a provably optimal detector candidate.

We focus on the scenario where DM dominantly couples to the electron density (as is the case for the benchmark model of a DM particle which interacts with the Standard Model via a kinetically-mixed dark photon) in a detector at zero temperature. 
As shown in Refs.~\cite{Hochberg:2021pkt,Knapen:2021run,Boyd:2022tcn}, for this interaction the target response can be written in terms of the electron energy-loss function, $\text{Im}\left[ -1 / \varepsilon(\mathbf{q}, \omega)\right]$, where $\varepsilon(\mathbf{q}, \omega)$ is the dielectric function.
This is especially useful in the context of limiting the DM-electron scattering rate since there are first-principles constraints on $\text{Im}\left[ -1 / \varepsilon(\mathbf{q}, \omega)\right]$.

These first-principles constraints are provided by well-known \emph{sum rules} from condensed matter physics~\cite{Mahan, Dressel_Gruner_2002},
\begin{align}
    \displaystyle\int_0^\infty \frac{\dd\omega}{\omega} \, \text{Im} \left[ \frac{-1}{\varepsilon(\mathbf{q}, \omega)} \right] &= \frac{\pi}{2} \left( 1 - \frac{1}{\varepsilon(\q, 0)} \right)\,, \label{eqn:sum_rule_1_w}\\
    \displaystyle\int_0^\infty {\dd\omega} \, \omega \, \text{Im} \left[ \frac{-1}{\varepsilon(\mathbf{q}, \omega)} \right] &= \frac{\pi}{2} \omega_\text{p}^2\, ,  \label{eqn:sum_rule_w}
\end{align}
where $\omega_\text{p}$ is the plasma frequency of the material. The first of these sum rules, Eq.~\eqref{eqn:sum_rule_1_w}, arises from causality and unitarity, since $\text{Im}\left[ -1 / \varepsilon(\mathbf{q}, \omega)\right]$ is a causal response function, while Eq.~\eqref{eqn:sum_rule_w} (known as the ``$f$-sum rule'') enforces charge conservation, since $\omega_\text{p}^2$ scales with the electron number density $n_e$. Note in particular that in general $n_e$ is \emph{not} just the valence electron density in a semiconductor, or the density of free charge carriers in a metal; rather, it counts \emph{all} electrons, including core electrons~\cite{pines1956collective}.

By combining Eqs.~\eqref{eqn:sum_rule_1_w} and~\eqref{eqn:sum_rule_w} using the Cauchy-Schwarz inequality, and its generalization the H\"{o}lder inequality, we derive a set of novel upper bounds on the DM-electron scattering rate. 
The use of Cauchy-Schwarz to derive a sum rule on $\text{Im}[-1/\varepsilon(\q,\omega)]$ without any weighting by powers of $\omega$ recently appeared in the condensed matter literature as a ``quantum weight sum rule''~\cite{PhysRevResearch.7.023158, Souza_2025, Verma_2025}. Similarly, Ref.~\cite{Lasenby:2021wsc} showed that one could obtain a strong upper bound on the DM-electron scattering rate from Eq.~\eqref{eqn:sum_rule_1_w} alone (see also Ref.~\cite{Catena:2025sxz}). Here, we show that under conservative assumptions, the combination of both sum rules yields a stronger bound for DM masses above the MeV scale. Additionally, we improve on the previous upper bounds from Ref.~\cite{Lasenby:2021wsc}, using only Eq.~\eqref{eqn:sum_rule_1_w}, by including realistic models for $\varepsilon(\q, 0)$, and discuss how a semi-analytic argument for the large-$q$ behavior of $\varepsilon(\q, 0)$ yields considerably stronger bounds for both heavy and light mediators.

This \textit{Letter} is organized as follows. We begin by deriving a family of upper bounds on the DM-electron scattering rate using the sum rules in Eqs.~\eqref{eqn:sum_rule_1_w} and~\eqref{eqn:sum_rule_w}.
While the most stringent upper bounds will depend on the details of the static dielectric function, $\varepsilon(\mathbf{q}, 0)$, we first derive new conservative bounds which depend only on $\omega_\text{p}$ and the target mass density.
We then derive improved bounds by incorporating a static dielectric function with more realistic momentum dependence.
This dielectric function is calculated with a combination of state-of-the-art density functional theory (DFT) tools and semi-analytic calculations, the details of which are discussed in the \textit{Supplemental Material}.
We conclude with some perspectives on the experimental program for sub-GeV DM searches in light of these new bounds.

\begin{figure*}[ht!]
    \centering
    \includegraphics[width=0.45\textwidth]{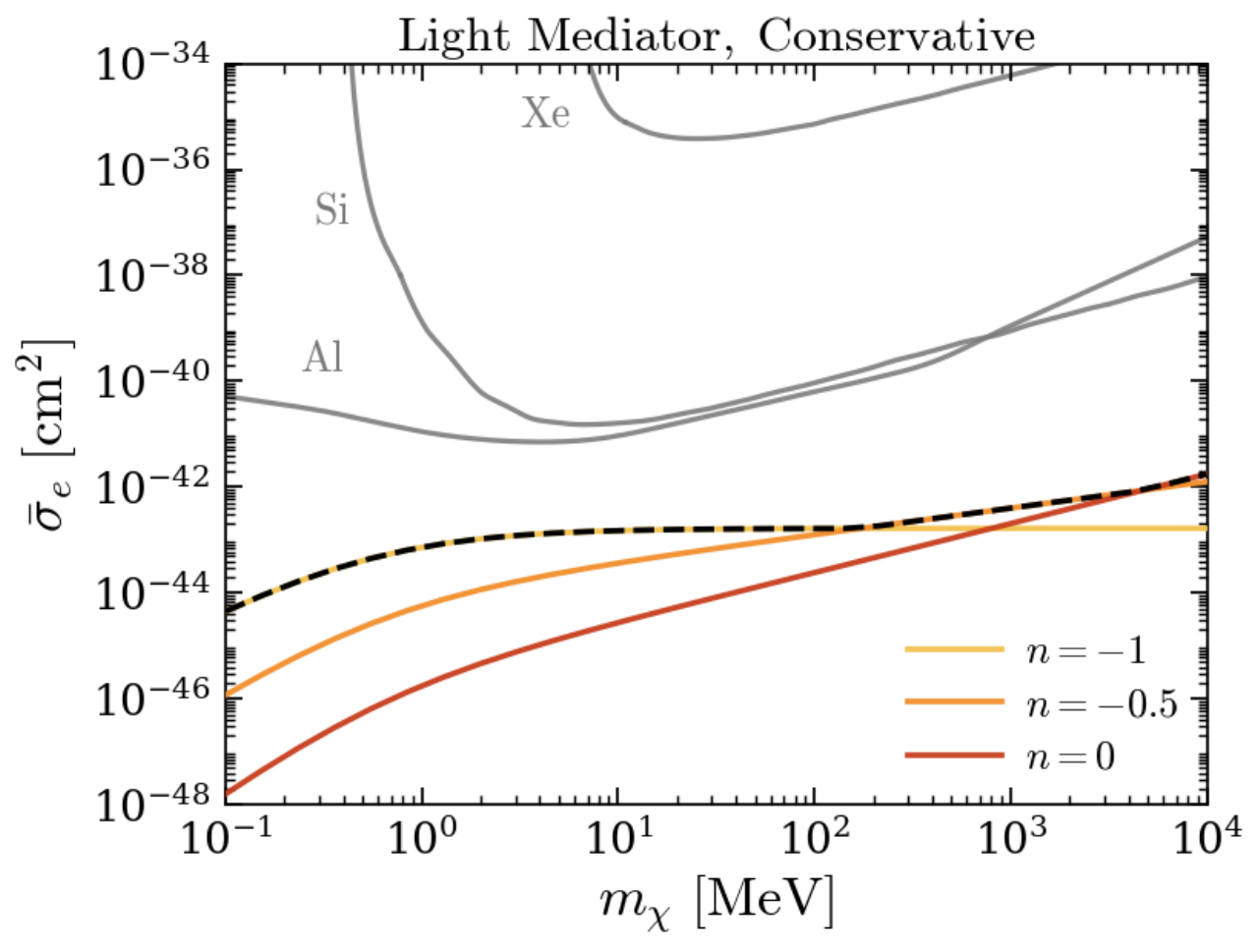}
    \hspace{1em}
    \includegraphics[width=0.45\textwidth]{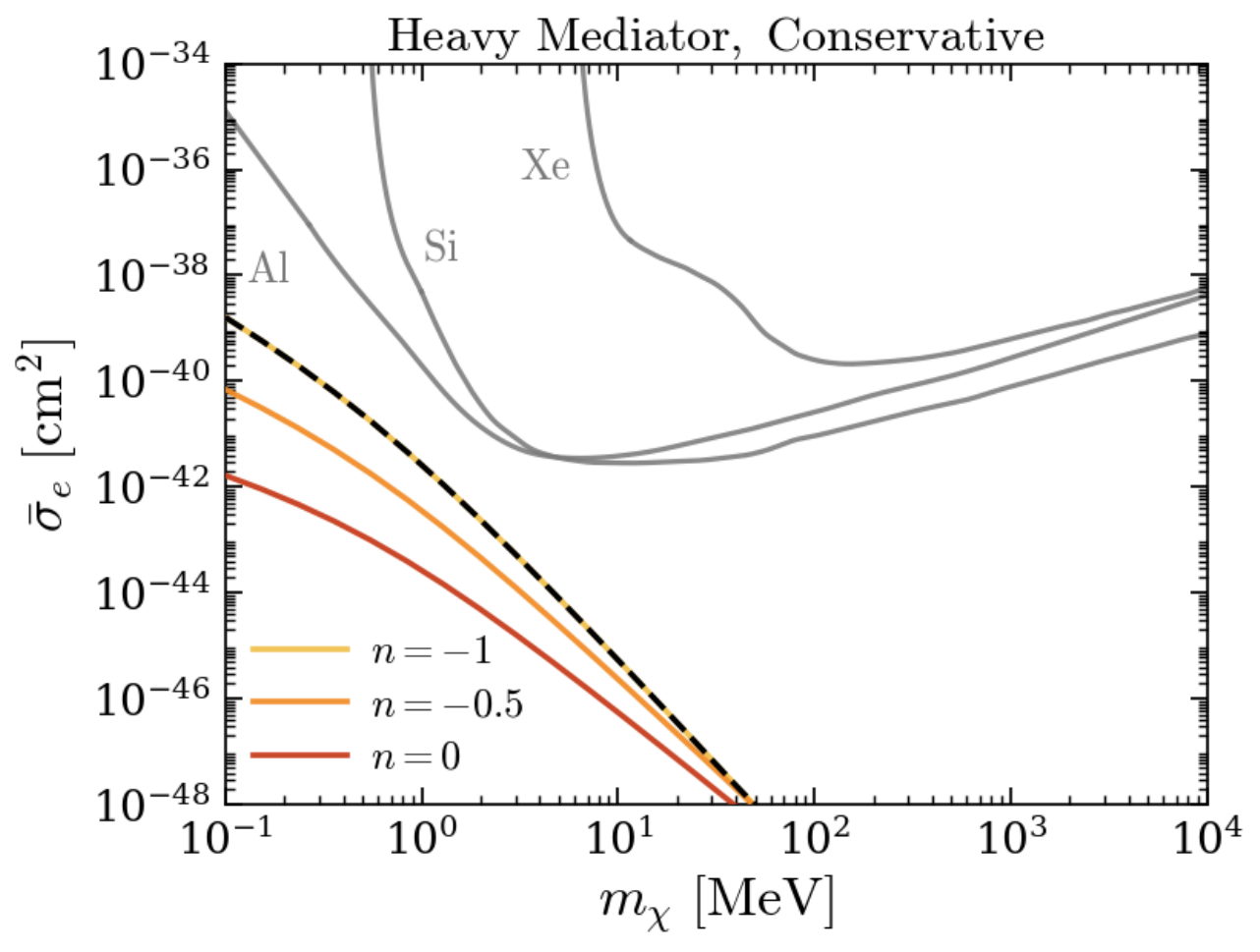}
    \caption{
   Comparison of the conservative lower bounds on the 95\% C.L. (3 events, zero background) cross section sensitivity $\bar{\sigma}_e$ of an experiment with a $1 \, \text{kg} \cdot \text{yr}$ exposure.
    The conservative bounds are computed using Eq.~\eqref{eq:final_rate_limit} for different $n$ with $1 - 1 / \varepsilon(q, 0) \rightarrow 1$, and shown as colored lines. 
    We assume $\omega_\text{p} = 30 \ \text{eV}$, corresponding to the approximate all-electron plasma frequency of solid silicon, aluminum and liquid xenon, and set $\omega_{\rm th} = 1 \ {\rm meV}$.
    The black dashed line corresponds to the outline of the most stringent limit at each $m_\chi$.
     The grey curves are the projected sensitivities for a liquid Xe~\cite{Pandey:2018esq} with a single-electron threshold, Si~\cite{Knapen:2021run} with a single-electron threshold, and Al~\cite{Hochberg:2021pkt} with a sensitivity of $10 \ {\rm meV} < \omega < 10 \ {\rm eV}$, all with a $1 \, \text{kg} \cdot \text{yr}$ exposure.}
    \label{fig:conservative_bounds}
\end{figure*}

\vspace{1em}
\noindent\textbf{Derivation of the upper bounds.}
If DM dominantly interacts with the electron number density, the DM-electron scattering rate is~\cite{Hochberg:2021pkt, Knapen:2021run}
\begin{align}
    R = \frac{\rho_\chi}{\rho_T m_\chi} \frac{\bar{\sigma}_e}{e^2 \mu_{\chi e}^2} \int_{\omega_{\rm th}}  \frac{\dd \omega \, \dd^3 \mathbf{q}}{(2\pi)^3} q^2 \mathcal{F}^2 \, g(\mathbf{q}, \omega) \text{Im} \left[ \frac{-1}{\varepsilon(\mathbf{q}, \omega)}\right] \, ,
    \label{eq:rate_general}
\end{align}
where $\omega_{\rm th}$ is the energy threshold of the experiment, $\rho_\chi \approx 0.4 \, \text{GeV} / \text{cm}^3$ is the DM density, $\rho_T$ is the target density, $\bar{\sigma}_e$ is the reference cross section, $e = \sqrt{4 \pi \alpha}$, where $\alpha$ is the fine-structure constant ($\alpha \approx 1 / 137$ in the Heaviside-Lorentz units used throughout), $\mu_{\chi e}$ is the DM-electron reduced mass, $\omega$ and $\mathbf{q}$ are the energy and momentum transferred to the target, respectively, and $q = |\mathbf{q}|$.
The mediator form factor, $\mathcal{F}$, determines the dependence on the mediating dark photon mass: $\mathcal{F}\approx 1$ in the heavy mediator limit, and $\mathcal{F}\approx (\alpha m_e /  q)^2$, where $m_e$ is the electron mass, in the light mediator limit; the applicability of these limits is defined precisely in Ref.~\cite{Stratman:2026qnh}. 
The kinematic function, $g(\mathbf{q}, \omega)$~\cite{Trickle:2019nya}, encodes the DM-electron scattering rate dependence on the DM velocity distribution, $f(\mathbf{v})$, through $g(\mathbf{q}, \omega) \equiv 2 \pi \int d^3 \mathbf{v} f(\mathbf{v}) \delta(\omega - \omega_\mathbf{q})$, where $\omega_{\mathbf{q}} = \mathbf{q} \cdot \mathbf{v} - q^2 / 2 m_\chi$ is the energy deposited to the target in a scattering event.
Lastly, the energy-loss function~\cite{PhysRev.113.1254}, $\text{Im}[-1 / \varepsilon(\mathbf{q}, \omega)]$, determines the target specific response, where $\varepsilon(\mathbf{q}, \omega)$ is the dielectric function
\footnote{$\varepsilon(\mathbf{q}, \omega)$ is the longitudinal dielectric when the target dielectric tensor is diagonal, which we assume throughout.}.

Equation~\eqref{eq:rate_general} can be further simplified for isotropic targets with $\varepsilon(\mathbf{q}, \omega) \approx \varepsilon(q, \omega)$,
\begin{align}
    R = \frac{\rho_\chi}{\rho_T m_\chi} \frac{\bar{\sigma}_e}{e^2 \mu_{\chi e}^2} \int \dd q \, q^3 \mathcal{F}^2 \, \int_{\omega_\text{th}} \frac{\dd \omega}{2 \pi} \eta(q, \omega) \, \text{Im} \left[ \frac{-1}{\varepsilon(q, \omega)}\right] \, ,
    \label{eq:rate_isotropic}
\end{align}
where $\eta(q, \omega) \equiv \int \dd^3 \mathbf{v} \, ( f(\mathbf{v}) / v ) \, \Theta(v - v_\text{min})$, and $v_\text{min} = \omega / q + q / 2 m_\chi$.
Note that a factor of $\omega^{-n} \, \eta(q, \omega)$ can be pulled out of the $\omega$ integral in Eq.~\eqref{eq:rate_isotropic} if we replace this factor with its maximum over all $\omega \geq \omega_\text{th}$. We define the function $\eta^\text{max}_n(q) \equiv \text{max}\{ \omega^{-n} \, \eta(q, \omega) \}$, which has the property that $\eta(q, \omega) \leq \omega^n \eta^\text{max}_n(q)$ for all $\omega$ and $n$.
Therefore the DM-electron scattering rate in Eq.~\eqref{eq:rate_isotropic} is bounded by,
\begin{align}
    R \le \frac{\rho_\chi}{\rho_T m_\chi} \frac{\bar{\sigma}_e}{2 \pi e^2 \mu_{\chi e}^2} \int \dd q \, q^3 \mathcal{F}^2 \, \eta^\text{max}_n(q) K_n(q) \, ,
    \label{eq:rate_isotropic_bound}
\end{align}
where 
\begin{align}
    K_n(q) \equiv \displaystyle\int_0^\infty \dd\omega \, \omega^n \, \text{Im}\left[ \frac{-1}{\varepsilon(q, \omega)} \right] \, .
    \label{eq:K_n}
\end{align}

We now use the H\"older inequalities to bound Eq.~\eqref{eq:K_n} with the sum rules in Eqs.~\eqref{eqn:sum_rule_1_w} and~\eqref{eqn:sum_rule_w}.
The H\"older inequalities~\cite{Holder1889, 1965hmfw.book.....A} state that for any two non-negative integrable functions, $f(\omega)$ and $g(\omega)$, and for any two real numbers $r,s \geq 1$ such that $1/r + 1/s = 1$, 
\begin{align}
    \displaystyle\int f(\omega) g(\omega) \, \dd \omega \le \left( \displaystyle\int [f(\omega)]^r \, \dd\omega \right)^{1/r} \left( \displaystyle\int [g(\omega)]^s \, \dd\omega \right)^{1/s} .
    \label{eq:Holder}
\end{align}
Setting $f(\omega)
=
\left(\omega\,\operatorname{Im}\!\left[
-\frac{1}{\varepsilon(q,\omega)}
\right]\right)^{1/r}$,~$g(\omega)
=
\left(\omega^{-1}\,\operatorname{Im}\!\left[
-\frac{1}{\varepsilon(q,\omega)}
\right]\right)^{1/s}$, and using Eqs.~\eqref{eqn:sum_rule_1_w} and~\eqref{eqn:sum_rule_w}, yields the following general family of inequalities,
\begin{align} 
    \label{eqn:Kn_bound}
    K_n(q) \le \frac{\pi}{2} \omega_\text{p}^{n+1} \left(\frac{ \varepsilon(q, 0) -1}{\varepsilon(q, 0)} \right)^{\frac{1 - n}{2}} \, ,
\end{align}
for $-1 \le n \le 1$. Substituting Eq.~\eqref{eqn:Kn_bound} into the DM-electron scattering rate in Eq.~\eqref{eq:rate_isotropic_bound} leads to our family of upper bounds on the DM-electron scattering rate, parameterized by $-1 \leq n \leq 1$, 
\begin{align}
    R \le \frac{\rho_\chi \bar{\sigma}_e}{\rho_T m_\chi} \frac{ \omega_\text{p}^{n + 1}}{4 e^2 \mu_{\chi e}^2}  \int \dd q \, q^3 \mathcal{F}^2 \, \eta^\text{max}_n(q) \left( \frac{\varepsilon(q, 0) - 1}{\varepsilon(q, 0)} \right)^\frac{1 - n}{2} \, .
    \label{eq:final_rate_limit}
\end{align}
This result generalizes the results of Ref.~\cite{Lasenby:2021wsc} which bounded the DM-electron scattering rate using only the sum rule in Eq.~\eqref{eqn:sum_rule_1_w}, which corresponds to the $n = -1$ limit of Eq.~\eqref{eq:final_rate_limit}.

\vspace{1em}
\noindent\textbf{Conservative cross section bounds.} The upper bounds on the DM-electron scattering rate in Eq.~\eqref{eq:final_rate_limit} correspond to lower bounds on the experimental sensitivity to $\bar{\sigma}_e$.
The tightest lower bounds will depend on the detailed electronic structure of the target via the static dielectric function, $\varepsilon(q, 0)$, which we will discuss in detail later.
However, more general, conservative, lower bounds on $\bar{\sigma}_e$ use the fact that the targets of interest (specifically, materials in their ground state at zero temperature) have $\varepsilon(q, 0) > 0$, and therefore $1 - 1 / \varepsilon(q, 0) \leq 1$.
Replacing $(1 - 1 / \varepsilon(q, 0)) \rightarrow 1$ in Eq.~\eqref{eq:final_rate_limit} then leads to lower bounds on $\bar{\sigma}_e$ that only depend on the target density, $\rho_T$, and plasma frequency, $\omega_\text{p}$.

In Fig.~\ref{fig:conservative_bounds} we show these conservative lower bounds for various $n$, assuming a light ($\mathcal{F} \approx (\alpha m_e / q)^2$, left panel) and heavy mediator ($\mathcal{F} \approx 1$, right panel), as different colored lines.
The dashed black line outlining the colored lines is the strongest conservative lower bound, corresponding to the most stringent bound across all $-1 \leq n \leq 1$ at each $m_\chi$. In both the light and heavy mediator limits, the most stringent lower bound on $\bar{\sigma}_e$ comes from $n = -1$ at small DM masses, with $n = 0$ becoming stronger at large DM masses. (Positive values of $n$ are strictly worse, so we do not show them for clarity.) All $\bar{\sigma}_e$ lower bounds correspond to the $95\%$ confidence level (C.L.) exclusion limits (3 events) assuming zero background and a 1 $\text{kg} \cdot \text{yr}$ exposure, using the recommended Standard Halo Model velocity parameters~\cite{Baxter:2021pqo} (velocity dispersion $v_0 = 238 \, \text{km} / \text{s}$, escape velocity $v_\text{esc} = 544 \, \text{km} / \text{s}$, and Earth galactic velocity $v_\text{e} = 250 \, \text{km} / \text{s}$) and taking a threshold energy of $\omega_\text{th} = 1 \, \mathrm{meV}$, at the expected scale of next-generation direct detection experiments~\cite{Kahn:2021ttr}.
Furthermore, for illustration we assume roughly material-agnostic values $\omega_\text{p} = 30 \, \text{eV}$ and $\rho_T = 3 \, \text{g} / \text{cm}^3$. The physical values for solid Al and Si and liquid Xe, when all electrons are accounted for, are $\omega_p^{\rm Al} = 32.9 \, {\rm eV}$, $\omega_p^{\rm Si} = 31.1 \, {\rm eV}$, $\omega_p^{\rm LXe} = 31.7 \, {\rm eV}$, and $\rho_T^{\rm Al} =  2.7 \, {\rm g/cm}^3$, $\rho_T^{\rm Si} =  2.3 \, {\rm g/cm}^3$, $\rho_T^{\rm LXe} = 2.9 \, {\rm g/cm}^3$. Theoretical calculations for the responses of these materials for various thresholds above our $\omega_\text{th}$ are shown in grey.

For light mediators (Fig.~\ref{fig:conservative_bounds}, left), the $m_\chi$ scaling of the lower bound on $\bar{\sigma}_e$ for $n = -1$ and $n = 0$ at large $m_\chi$, can be understood analytically.
For $n = -1$, $\eta^\text{max}_{-1}(q) \propto q \, \Theta(q_\text{max} - q)$, where $q_\text{max} = 2 m_\chi v_\text{max}$ is the maximum momentum transfer to the target and $v_\text{max} = v_\text{esc} + v_\text{e}$ is the maximum DM velocity in the lab frame.
This implies that the $q$ integral in Eq.~\eqref{eq:final_rate_limit}, in the conservative limit, scales linearly with $m_\chi$, leading to an $m_\chi$-independent limit on $\bar{\sigma}_e$, as was previously found in Ref.~\cite{Lasenby:2021wsc}.
However for $n = 0$, $\eta^\text{max}_0(q) \propto 1$ for $q_\text{min} \lesssim q \lesssim q_\text{max}$, where $q_\text{min} = \omega_\text{th} / v_\text{max}$, and therefore the $q$ integral in Eq.~\eqref{eq:final_rate_limit} only depends logarithmically on $m_\chi$ and $\omega_\text{th}$. This leads to the DM-electron scattering rate scaling as $R \propto \bar{\sigma}_e \log(q_\text{max} / q_\text{min}) / m_\chi$, and therefore the lower bound on $\bar{\sigma}_e \propto m_\chi / \log(q_\text{max} / q_\text{min})$.
We find the $n = -0.5$ limit starts to dominate around $m_\chi \gtrsim 100 \, {\rm MeV}$, and $n = 0$ becomes dominant for $m_\chi \gtrsim 5 \, \text{GeV}$.

The $m_\chi$ scaling of the conservative lower bounds on $\bar{\sigma}_e$ is dramatically different for heavy mediators (Fig.~\ref{fig:conservative_bounds}, right).
For $n = -1$, $R \propto \bar{\sigma}_e m_\chi^4$, and therefore the conservative lower bound on $\bar{\sigma}_e$ scales as $1/m_\chi^4$, while for $n = 0$  $R \propto \bar{\sigma}_e m_\chi^3 $ and $\bar{\sigma}_e \propto 1 / m_\chi^3$.
Therefore, as in the light-mediator case, the $n = 0$ limit eventually dominates over the $n = -1$ limit for larger $m_\chi$, though this occurs below the range of the $y$-axis in Fig.~\ref{fig:conservative_bounds}, right.
Relative to the light mediator scenario, the conservative limits of the heavy mediator scenario are not particularly constraining: all currently-operating detectors are several orders of magnitude above the bounds for DM masses well above the detector thresholds.
However, these limits are dominated by large $q$, where we expect the dielectric factor in Eq.~\eqref{eq:final_rate_limit} to cut off the $q$ integral before the kinematic boundary, and thus we expect considerably tighter bounds for realistic material responses. We turn to this situation next.

\begin{figure}[t!]
    \centering
    \includegraphics[width=\linewidth]{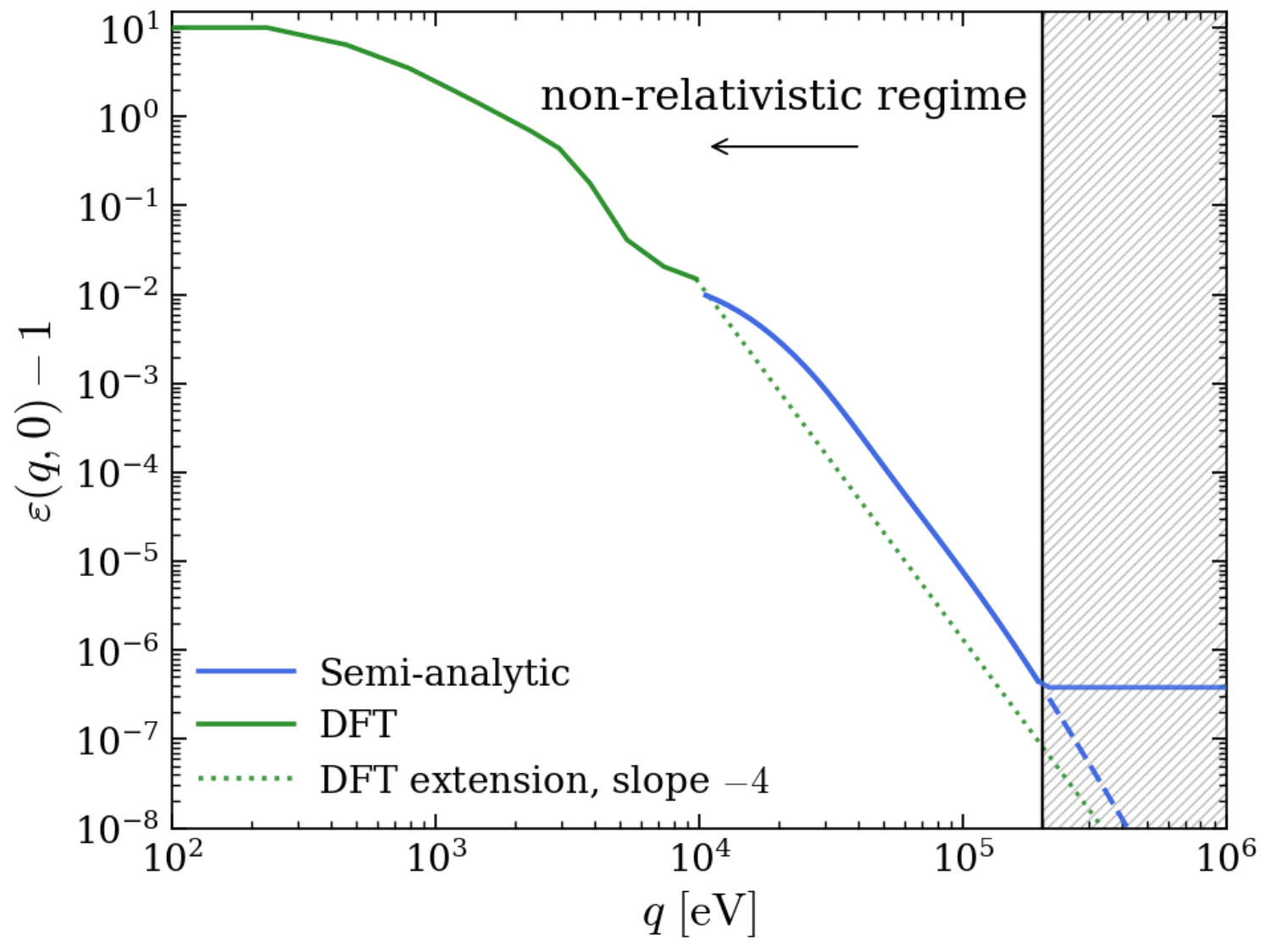}
    \caption{Static dielectric response of silicon, $\varepsilon(q,0)-1$. Blue: $\varepsilon(q,0)-1$ calculated from the semi-analytic model, set to a constant beyond the non-relativistic regime (grey hashed region) as a conservative approximation. Dashed blue indicates the result of the semi-analytic model outside the non-relativistic regime of validity. Green: the directionally-averaged DFT data, extended with a $q^{-4}$ extrapolation (dotted) to show the qualitative and quantitative agreement with the atomic model.}    
  \label{fig:epsilon_m1}
\end{figure}

\vspace{1em}
\noindent\textbf{Improved cross section bounds.} 
An improved upper bound on the DM-electron scattering rate can be achieved by restoring $\varepsilon(q, 0)$ in Eq.~\eqref{eq:final_rate_limit}. The static dielectric function at zero temperature is given by the Lindhard formula~\cite{Dressel_Gruner_2002,Mitridate:2021ctr,Krnjaic:2024bdd},
\begin{align}
    \varepsilon(q, 0) & = 1 - \frac{e^2}{V q^2} \sum_{J,J'} \frac{f_{J'} - f_J}{E_{J'} - E_J} \left| \langle J' |  e^{i \mathbf{q} \cdot \mathbf{x}}| J \rangle\right|^2 \, ,
    \nonumber \\
    & = 1 + \frac{2e^2}{Vq^2} \sum_{I,F} \frac{\left| \langle F |  e^{i \mathbf{q} \cdot \mathbf{x}}| I \rangle\right|^2}{E_F - E_I} \, ,
    \label{eq:static_dielectric_simple}
\end{align}
where $V$ is the target volume, $J$ indexes all the electronic states in the system, $| J \rangle$ are the (dimensionless) electronic states with energy levels $E_J$, and $f_J$ are the zero-temperature Fermi occupation factors: $f_J = 1 \, (0)$ for a filled (un-filled) electronic state.
The second line of Eq.~\eqref{eq:static_dielectric_simple} is a simplification of the first; $I \,(F)$ run over the filled (un-filled) electronic states, and we have assumed parity is a good symmetry of the electronic Hamiltonian so that $\left| \langle F |  e^{-i \mathbf{q} \cdot \mathbf{x}}| I \rangle\right|^2 = \left| \langle F |  e^{i \mathbf{q} \cdot \mathbf{x}}| I \rangle\right|^2$.

The calculation of the static dielectric for $\text{keV} \lesssim q \lesssim \text{MeV}$ in a real material is complicated and target-dependent due to the variety of electronic transitions that are included in Eq.~\eqref{eq:static_dielectric_simple}.
Using Si as an illustrative example, we employ a combination of numeric and analytic approaches to approximate $\varepsilon(q, 0)$ which we discuss in detail below and are summarized in Fig.~\ref{fig:epsilon_m1}.

\textit{Density-functional theory.} At $q \lesssim 10 \, \text{keV}$, $\varepsilon(q,0)$ will be dominated by valence-to-conduction transitions. 
We compute $\varepsilon(q,0)$ in silicon up to $q = 11.8 \, {\rm keV}$ using first-principles DFT calculations performed with the \textsf{Quantum Espresso} package~\cite{Gianozzi2009, Gianozzi2017, Gianozzi2020} with a plane-wave basis in the generalized-gradient approximation (GGA) as implemented by Perdew, Burke, and Ernzerhof (PBE)~\cite{Perdew1996}. The electronic structure was calculated using a 100 Ry energy cut-off on a $10 \times 10 \times 10$ $k$-grid with a scalar relativistic norm-conserving pseudopotential including 4 valence electrons per Si atom from the Pseudo-Dojo project~\cite{Hamann2013}. The dielectric function was calculated using the \textsf{BerkeleyGW} package~\cite{Hybertsen1986, Deslippe2012} with a scissor correction applied to the DFT eigenvalues to match the experimental band gap of Si. We find $\varepsilon(q,0) \simeq \mathcal{O}(1)$ for $q \lesssim 1 \ {\rm keV}$ and our calculations agree with the measured value $\varepsilon(0, 0) \approx 11.3$~\cite{PhysRevB.47.9892} at the smallest $q$; we expect $\varepsilon(q,0) \simeq \mathcal{O}(1)$ for most materials for momentum transfers small compared to the first Brillouin zone (1BZ). The directionally-averaged $\epsilon(q,0)-1$ from DFT is shown in green in Fig.~\ref{fig:epsilon_m1}.

\begin{figure*}[t!]
    \centering
    \includegraphics[width=0.45\textwidth]{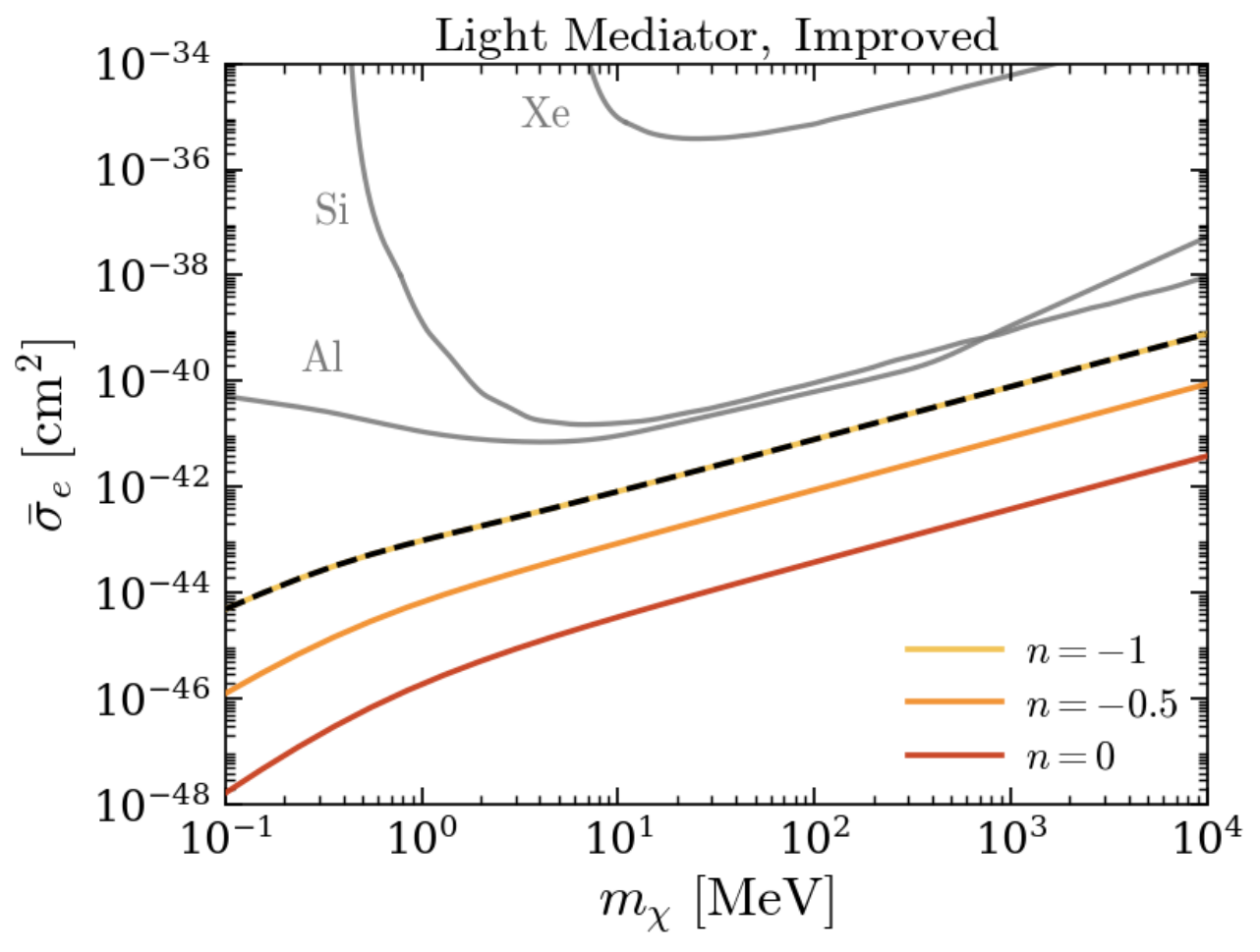}
    \hspace{1em}
    \includegraphics[width=0.45\textwidth]{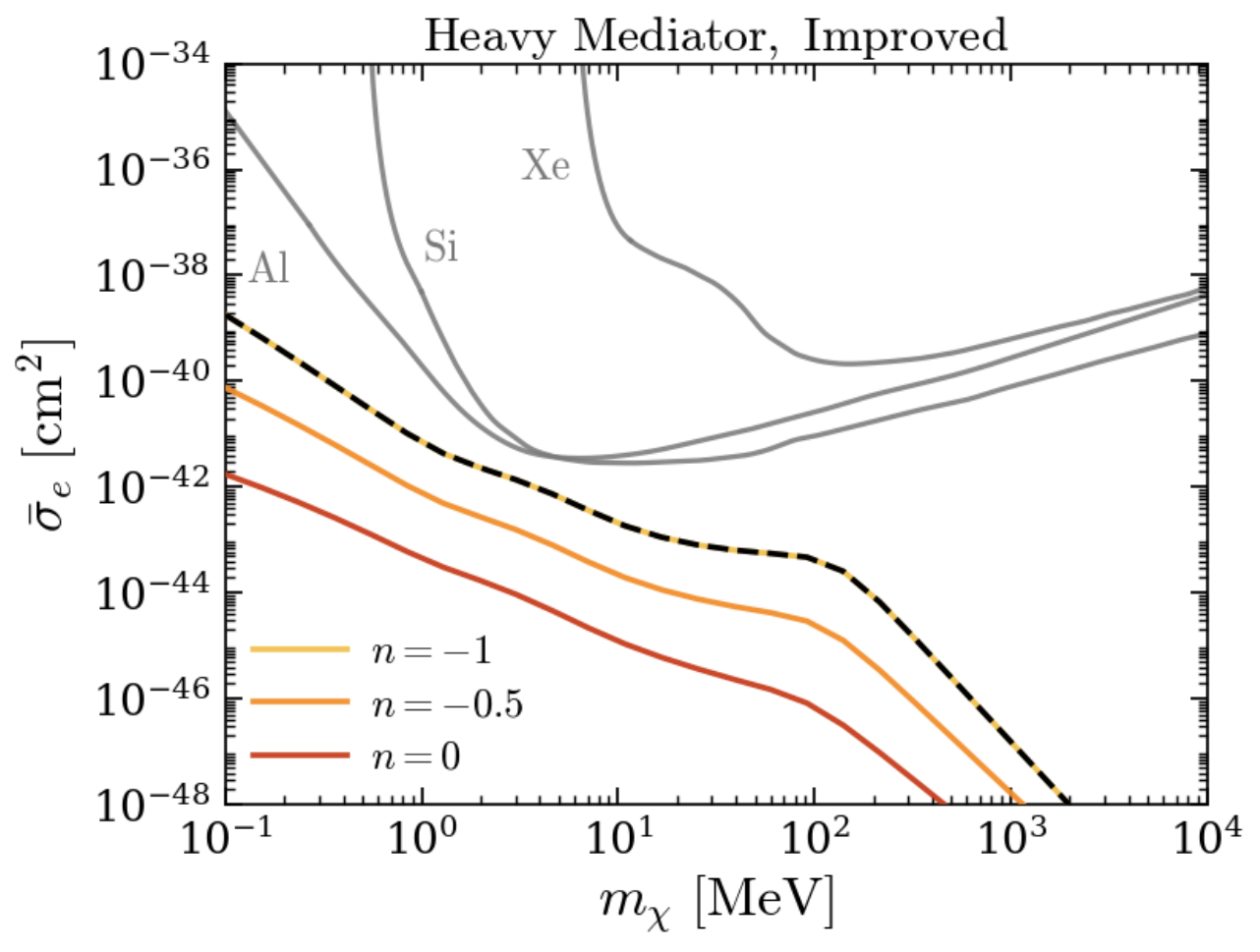}
    \caption{
    Comparison of the improved lower bounds on the 95\% C.L. (3 events, no background) cross section sensitivity of an experiment with a $\text{kg} \cdot \text{yr}$ exposure.
    The improved lower bounds, for different $n$, are computed using Eq.~\eqref{eq:final_rate_limit} with the static dielectric shown in Fig.~\ref{fig:epsilon_m1} (see text for details). All other parameters and curves are the same as in Fig.~\ref{fig:conservative_bounds}.}
    \label{fig:improved_limits}
\end{figure*}

\textit{Semi-analytic core-to-free.} For $q \gtrsim 10 \, \text{keV}$, the static dielectric begins to gain contributions from the core electronic states. 
To understand this contribution we model the initial, core electron states (1$s$, 2$s$, and 2$p$) with Roothaan--Hartree--Fock (RHF) wave functions~\cite{RevModPhys.32.179}, which are a linear combination of atomic orbitals tuned to solve the time-independent Schr\"odinger equation of an isolated atom.
The final states are modeled as Coulomb plane waves; a detailed discussion of the form of the electronic wavefunctions assumed, and the calculation of the transition matrix elements can be found in the \textit{Supplemental Material}
\footnote{This approach for modeling the core-to-free transitions is similar to that used in Ref.~\cite{Griffin:2021znd}. We omit the corresponding ``valence-to-free'' and ``core-to-conduction'' contributions since we are primarily focused on the large $q$ asymptotic behavior, for which we expect core-to-free transitions to dominate.}. The result is shown in blue in Fig.~\ref{fig:epsilon_m1}, restricting to momentum transfers $q \gtrsim 10 \ {\rm keV}$ where the bound electron response is expected to dominate. At momenta $q$ such that $q^2/(2m_e) \gtrsim 0.1 m_e$ (grey hashed region in Fig.~\ref{fig:epsilon_m1}), the final-state energies are relativistic and the non-relativistic Lindhard formula no longer applies (blue dashed). Since larger $\varepsilon$ suppresses the rate integral less, we conservatively account for our ignorance in the large-$q$ regime by setting $\varepsilon$ to a constant value equal to its numerical value in the semi-analytic approach at $q = 200 \ {\rm keV}$.

\textit{Analytic approximation.} In addition to the semi-analytic approach to understanding the large $q$ dependence of the static dielectric, there is a useful analytic limit of Eq.~\eqref{eq:static_dielectric_simple}.
If the final electronic states are well-described by plane waves, the static dielectric can be approximated as,
\begin{align}
    \varepsilon(q, 0) \approx 1 + \frac{4 m_e e^2}{V \, q^2}\sum_I \int \frac{\dd^3 \mathbf{k}}{(2\pi)^3}\frac{| \widetilde{\psi}_I(\mathbf{k})|^2}{(\mathbf{k}-\mathbf{q})^2-2 m_eE_I} \, ,
    \label{eq:static_dielectric_analytic_approx_1}
\end{align}
where $\tilde{\psi}_I(\mathbf{k})$ is the Fourier transform of the initial state wavefunction.
Furthermore, if $q$ is much larger than both $\sqrt{2 m_e |E_I|}$ and the Fourier components $\mathbf{k}$ where the initial-state wavefunction is large, then Eq.~\eqref{eq:static_dielectric_analytic_approx_1} simplifies further. Since the initial-state wavefunction is normalized to 1 for a single electron, the sum over initial states and the volume factor combine to give the total electron density $n_e$, and we obtain
\begin{align}
    \varepsilon(q, 0) \approx 1 + \frac{4 m_e^2 \omega_\text{p}^2}{q^4} \, .
    \label{eq:static_dielectric_asymptotic}
\end{align}
This matches the large $q$ scaling of other analytic approximations of the dielectric in semiconductor targets~\cite{PhysRevB.47.9892}, and is expected to be a good approximation for electronic states where the assumptions are satisfied, i.e., when $q \gg Z_\text{eff} \alpha m_e \sim Z_\text{eff} \, (3.7 \, \text{keV})$, where $Z_\text{eff}$ is an effective nuclear charge and $q \gg \sqrt{2 m_e |E_I|} \sim 10 \, \text{keV} \, \left( |E_I| / 100 \, \text{eV} \right)^{1/2}$.
This corresponds to roughly $10 \, \text{keV} \lesssim q \lesssim 100 \, \text{keV}$ for weakly-bound states, in the intermediate range between valence and core, where we find good agreement with the semi-analytic solution shown in Fig.~\ref{fig:epsilon_m1}. Indeed, a simple extrapolation of the DFT result scaling as $q^{-4}$ (green dotted) is within an order of magnitude of our atomic orbital calculation.

\textit{Improved sensitivity bounds.} We include the momentum-dependent material response from Fig.~\ref{fig:epsilon_m1} in Eq.~\eqref{eq:final_rate_limit} by using the DFT result for $q \leq 11.8 \ {\rm keV}$ and the semi-analytic result for larger $q$, with $\varepsilon(q,0) - 1$ set to its constant value for $q \geq 200 \ {\rm keV}$. (The small discontinuity between the DFT and semi-analytic results does not qualitatively affect our results because the dielectric factor only appears in an integral over all $q$.) Doing so, we obtain the bounds on $\bar{\sigma}_e$ shown in Fig.~\ref{fig:improved_limits}. The impact of the dielectric is most transparent in the light-mediator case. The integrand contains the factor $\left(1-\frac{1}{\varepsilon(q,0)}\right)^{(1-n)/2} \propto q^{2n-2}$, which multiplies the rate integrand by negative powers of $q$ and weights the integral more toward small $q$. This suppresses the integral for large $m_\chi$. The effect is strongest for $n = -1$, which becomes the strongest bound for both the light and heavy mediators up to a DM mass of 10 GeV. While naively this would seem to make our tightest bound independent of $\omega_\text{p}$ because the $\omega_\text{p}$ prefactors in Eq.~\eqref{eq:final_rate_limit} cancel, the asymptotic scaling of the static dielectric does depend on $\omega_\text{p}$ through Eq.~\eqref{eq:static_dielectric_asymptotic}. We see that the total electron response of the material controls the overall scale of the bounds just as in the conservative case, though there is some material dependence through the specific shape of the static dielectric at both small (1BZ) and large (relativistic) $q$.

\vspace{1em}
\noindent\textbf{Conclusions.} 
In this \textit{Letter}, we have derived general bounds on the sensitivity of direct-detection experiments searching for DM-electron scattering through the coupling to electron density. These bounds form a continuous family obtained from electromagnetic sum rules that follow from first principles, including causality and charge conservation. They provide fundamental limits on DM-electron scattering experiments while reducing the dependence on material properties to the static dielectric function, $\varepsilon(\q,0)$, together with two bulk quantities: the target density, $\rho_T$, and plasma frequency, $\omega_{\rm p}$. Using the Lindhard formalism and DFT calculations, we find that the static dielectric function exhibits a generic power-law scaling at large momentum transfer: $\varepsilon(q, 0) - 1\propto q^{-4}$. In this regime, its behavior is largely determined by $\omega_{\rm p}$, so that the dielectric response is not an independent material parameter, and the upper bounds on the rate only depend on $\rho_T$ and $\omega_{\rm p}$. These bounds therefore provide both a useful consistency check as well as a theoretical benchmark for evaluating proposed materials for DM direct detection. Indeed, given that conventional materials such as silicon and aluminum are already within an order of magnitude of the improved bound for light mediators over the entire 10 MeV -- 10 GeV range of DM masses (and within an even smaller $\mathcal{O}(1)$ factor for DM masses around 5 MeV with a heavy mediator), the benefits of searching for more bespoke materials in this part of parameter space may be outweighed by simply scaling up the target mass of the conventional materials. That said, our results are derived assuming zero background; in the presence of a large background rate, the reach of even an optimal material will saturate, while an anisotropic material with daily modulation sensitivity has a reach that continues to improve with exposure~\cite{Blanco:2026kda}.

Our bounds can also be derived for different assumptions about the DM phase space distribution. Although the bounds presented here are evaluated using the Standard Halo Model, the derivation applies equally well to arbitrary velocity distributions. The same approach can therefore be applied to nonstandard populations, including tidal streams~\cite{Freese:2003na, Freese:2003tt, OHare:2014nxd}, cosmic ray-boosted DM~\cite{Bringmann:2018cvk}, and solar-reflected DM~\cite{An:2017ojc, An:2021qdl}. More broadly, these bounds provide a largely target-independent reference point for the sub-GeV direct-detection program and quantify how closely existing and proposed experiments approach the maximum scattering rates allowed by fundamental physical principles. 

\textit{Acknowledgments.} We are indebted to Peter Abbamonte for bringing the quantum weight sum rule to our attention, and for emphasizing the importance of the $f$-sum rule for dark matter detection. We thank Sin\'{e}ad Griffin and Omar Ashour for enlightening conversations regarding DFT computations of dielectric functions. 
B.A.B. acknowledges support from the National Science Foundation Award ID No. 2427159. L.J. is supported by the Connaught Fund of the University of Toronto. Y.K. acknowledges the support of a Discovery Grant from the
Natural Sciences and Engineering Research Council of Canada (NSERC). %
E.A.P. acknowledges funding via Los Alamos National Laboratory, via the U.S. DOE NNSA
under Contract No. 89233218CNA000001, through the LANL LDRD program, project number 20220135DR. E.A.P. acknowledges computational resources provided in part by the Center
for Integrated Nanotechnologies, a DOE Office of Science user facility, in partnership with the LANL Institutional Computing Program. E.A.P. performed additional calculations at the National Energy Research Scientific Computing Center (NERSC), a U.S. Department of Energy Office of Science User Facility located at Lawrence Berkeley National Laboratory, under NERSC award ERCAP0020494. %
A.P. acknowledges support from the Leinweber Foundation and from DOE grant DE-SC0025293.
T.T. is supported by the DOE grant DE-SC0015655.
S.L.W. acknowledges support from the LANL Director’s Postdoctoral Fellowship award 20230782PRD1.
This work used Expanse at the San Diego Supercomputer Cluster at UC San Diego through allocation PHY250392 from the Advanced Cyberinfrastructure Coordination Ecosystem: Services \& Support (ACCESS) program, which is supported by U.S. National Science Foundation grants \#2138259, \#2138286, \#2138307, \#2137603, and \#2138296.

\bibliographystyle{utphys}
\bibliography{biblio}


\appendix
\onecolumngrid

\newpage 

\renewcommand{\theequation}{S.\arabic{equation}}
\renewcommand{\thefigure}{S.\arabic{figure}}
\setcounter{equation}{0}
\setcounter{figure}{0}

\section{\large Supplemental Material: Semi-Analytical Calculation of the Static Dielectric Response}

\vspace{-0.5em}
\begin{center}
    Bradford A. Barker, Jay Epstein, Luke James, Yonatan Kahn, Elizabeth A. Peterson, Anirudh Prabhu, Tanner Trickle, and Samuel L. Watkins
\end{center}

In this Supplemental Material, we describe the semi-analytical calculation of the static dielectric response used in the main text.  As in the main text, we work in Heaviside-Lorentz units, so \(e^2=4\pi\alpha\), and the Coulomb prefactor is written as $e^2/q^2$.\\

We model the occupied initial states $| I\rangle$ in Eq.~\eqref{eq:static_dielectric_simple} as orbitals of the isolated neutral silicon atom described by Roothaan-Hartree-Fock (RHF) wavefunctions. For an occupied shell with principal and orbital quantum numbers $(n,l)$, the radial wavefunction is written as a linear combination of Slater-type orbitals,
\begin{align}
    R_{nl}(r)=\sum_j C_j\frac{(2Z_j)^{n_j+1/2}}{\sqrt{(2n_j)!}}r^{n_j-1}
    e^{-Z_j r}.
    \label{eq:app_RHF_STO}
\end{align}
The coefficients $C_j$, exponents $Z_j$, and effective principal quantum numbers $n_j$ are taken from tabulated RHF calculations~\cite{Bunge:1993rhf}. The reduced radial function is $u_{nl}(r)=rR_{nl}(r)$, normalized according to $\int_0^\infty dr |u_{nl}(r)|^2 = 1$.\\

The outgoing electron state, $| F \rangle$, is described by a continuum Coulomb wave that is an eigenstate of the attractive Coulomb potential, $V(r) = -Z_{\rm eff} / r$, where $Z_\text{eff}$ is an effective nuclear charge. Following the hydrogen-like continuum approximation of Ref.~\cite{Catena:2020AtomicResponses}, we assign each
shell a value of $Z_{\mathrm{eff}}$ fixed by its RHF binding energy,
\begin{align}
    Z_{\rm eff}(n,l)=n\sqrt{-2E_{nl}},
    \label{eq:app_Zeff}
\end{align}
with energies expressed in atomic units. 
The reduced continuum radial wavefunction may be written in terms of the regular Coulomb function as
\begin{align}
    u_{k l}(r)=\sqrt{\frac{2}{\pi}}F_{l}(\eta,kr),
    \qquad \eta=-\frac{Z_{\rm eff}}{k},
    \label{eq:app_coulomb_continuum}
\end{align}
For fixed $\ell$ and $Z_{\mathrm{eff}}$, these continuum functions are
normalized with respect to the asymptotic momentum according to
\begin{equation}
    \int_0^\infty dr\,
    u_{k\ell}^{*}(r)u_{k'\ell}(r)
    =
    \delta(k-k').
\end{equation}
We use this momentum-normalized convention throughout, so that the
continuum sum over final states in each partial wave carries the measure
$\int_0^\infty dk$.\\

\begin{figure}[ht!]
    \centering
    \includegraphics[width= 0.5\linewidth]{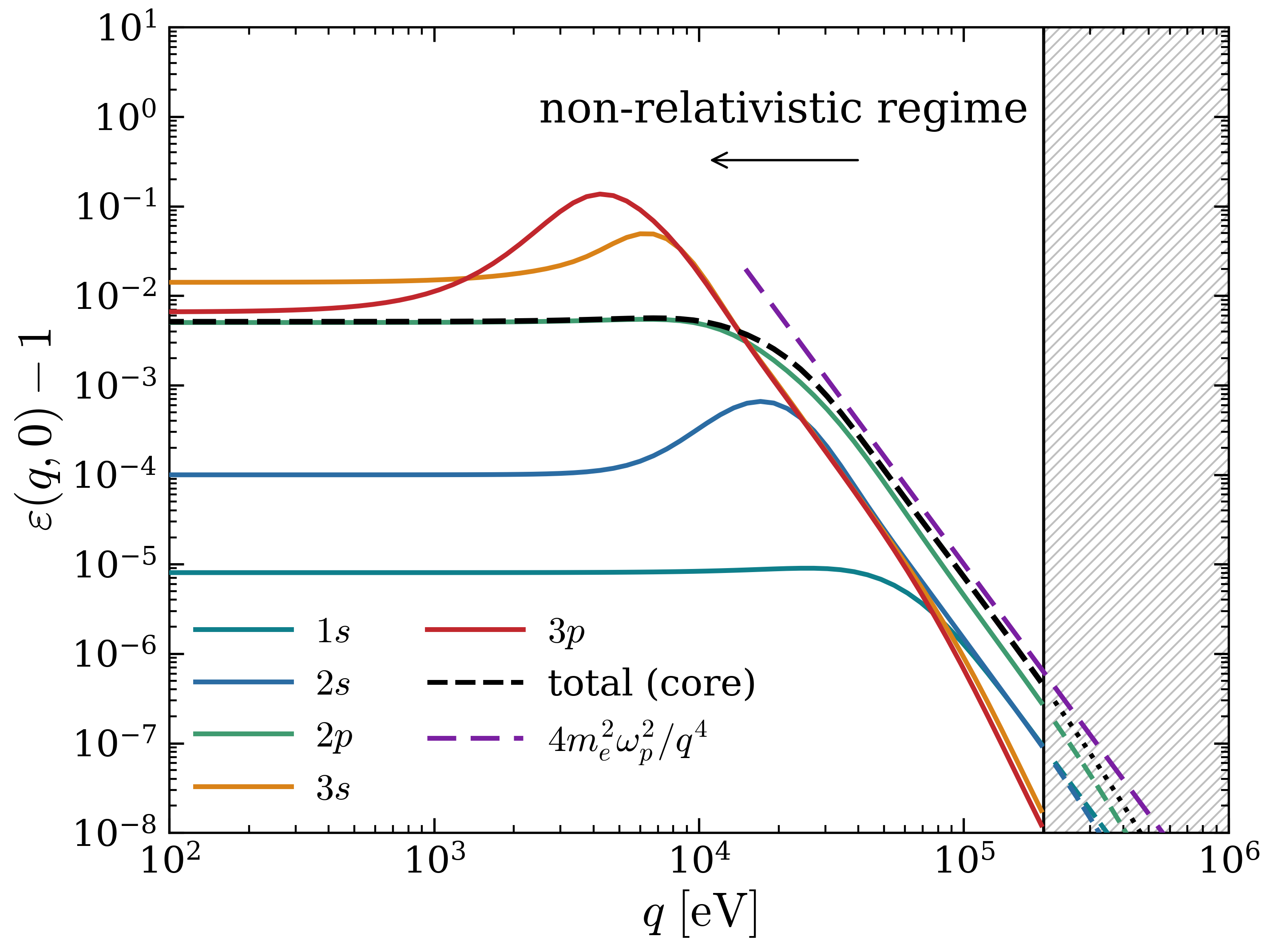}
    \caption{Contributions to the static dielectric function of silicon in the atomic model, broken out by initial-state orbital. The sum of the core shells (excluding $3s$ and $3p$), shown in dashed black, has a large-$q$ slope which closely matches the $q^{-4}$ expectation from the semi-analytic calculation in the main text, Eq.~(\ref{eq:static_dielectric_asymptotic}).}
  \label{fig:shell_resolv}
\end{figure}

Given these initial and final state approximations, we now turn to computing the basic transition matrix element appearing in Eq.~\eqref{eq:static_dielectric_simple},
\begin{align}
M_{IF}(\mathbf{q})
=
\langle F|e^{i\mathbf{q}\cdot\mathbf{x}}|I\rangle.
\end{align}
Writing the initial and final states as
\begin{align}
    |I\rangle = |n_i l_i m_i s_i\rangle,
    \qquad
    |F\rangle = |k l_f m_f s_f\rangle,
\end{align}
and using the spin-independence of the density operator, the matrix element factorizes as
\begin{align}
    \mathcal{M}_{IF}(\mathbf{q})
    =
    M_{if}(\mathbf{q})\delta_{s_fs_i}.
\end{align}
We then expand the plane wave as
\begin{align}
    e^{i\mathbf{q}\cdot\mathbf{r}}=4\pi\sum_{L,M}i^Lj_L(qr)Y^*_{LM}(\hat{\mathbf{q}})Y_{LM}(\hat{\mathbf{r}}).
    \label{eq:app_plane_wave}
\end{align}
This gives
\begin{align}
    M_{if}(k,\textbf{q})
    =
    4\pi
    \sum_{L,M}
    i^L Y_{LM}^{*}(\hat{\mathbf q})
    I_L(k,q)G_{if}^{LM}.
\end{align}
where
\begin{align}
    I_L(k,q)
    =
    \int_0^\infty dr\,r^2
    R_{k l_f}(r)R_{n_i l_i}(r)j_L(qr)
    \label{eq:radial-integral}
\end{align}
is the radial transition integral, and $G^{LM}_{if}$ is the Gaunt coefficient, and $u_{kl} = r R_{kl}$. In terms of Wigner-3j symbols,
\begin{align}
    G^{LM}_{if}=(-1)^{m_f}\sqrt{\frac{(2l_f+1)(2L+1)(2l_i+1)}{4\pi}}
    \begin{pmatrix}
        l_f & L & l_i\\
        0 & 0 & 0
    \end{pmatrix}
    \begin{pmatrix}
        l_f & L & l_i\\
        -m_f & M & m_i
    \end{pmatrix}.
    \label{eq:app_gaunt_3j}
\end{align}
 The allowed values of the multipoles $L$ are those satisfying the triangle rule and parity condition,
\begin{align}
    |l_i-l_f|\leq L\leq l_i+l_f,
    \qquad
    l_i+l_f+L \ \text{even}.
    \label{eq:app_allowed_L}
\end{align}
Because we are interested in the isotropic response, we average over the direction of \(\mathbf{q}\) and sum over magnetic quantum numbers. Using
\begin{align}
    \int d\Omega_q
    Y^*_{LM}(\hat{\mathbf{q}})
    Y_{L'M'}(\hat{\mathbf{q}})
    =
    \delta_{LL'}\delta_{MM'},
\end{align}
and the orthogonality properties of the Wigner-3j symbols, one obtains
\begin{align}
    \frac{1}{4\pi}\int d\Omega_q\sum_{m_i,m_f}|M_{if}(\mathbf{q})|^2
    =
    \sum_Lw(l_i,L,l_f)|I_L(k,q)|^2,
    \label{eq:app_angavg_result}
\end{align}
where
\begin{align}
    w(l_i,L,l_f)
    =
    (2l_i+1)(2L+1)(2l_f+1)
    \begin{pmatrix}
        l_i & L & l_f\\
        0 & 0 & 0
    \end{pmatrix}^2.
    \label{eq:app_angular_weight}
\end{align}
The factors of $4\pi$ from the plane-wave expansion, the angular average, and the Gaunt-coefficient normalization all cancel in Eq.~\eqref{eq:app_angavg_result}. Thus the radial integral $I_L(k,q)$ in Eq.~\eqref{eq:radial-integral} carries no additional factor of $4\pi$. The static dielectric response thus reads:
\begin{align}
    \varepsilon(q,0) - 1 &= \sum_{(n,\,l_i)} \sum_{l_f} \varepsilon_{if}(q,0),
\label{eq:channel-sum}\\[4pt]
\varepsilon_{if}(q,0) &= \frac{2\, e^{2}\, n_{\mathrm{at}}}{q^{2}}
\cdot 2 f_{\mathrm{occ}}
\sum_{L} \int_{0}^{\infty} dk\;
\frac{\; w(l_i, L, l_f)\, \big|I_L(k,q)\big|^{2}}{E_f(k) - E_{nl}},
\label{eq:master}
\end{align}
where the first factor of 2 is the anti-resonant factor arising from the two
time orderings in the static limit,
the second factor of 2 is the spin sum, and
$f_{\mathrm{occ}}$ is the fractional occupation of the shell. 
The number density $n_{\mathrm{at}}$ is obtained by summing over the $N$ scattering centers in the target.

The continuum integration of equation~\eqref{eq:master} is performed on a logarithmic grid in the final state momentum $k$. The integration window is defined in terms of the energy transfer $\omega = E_f(k) - E_{nl}$, with
$\omega_{\mathrm{min}} \leq \omega \leq \omega_{\mathrm{max}}$ taken
common to all channels. Because the initial-state RHF wavefunctions are not orthogonal to the final-state Coulomb wavefunctions given our prescription for $Z_{\rm eff}$, we neglect the $l_i = l_f$ contributions which would give a spurious nonzero matrix element $\mathcal{M}_{IF}(\mathbf{q})$ which does not vanish as $q \to 0$. The $l_i \neq l_f$ matrix elements are guaranteed to vanish as positive powers of $q$ by the orthogonality of the spherical harmonics. Summing over all occupied shells then yields the full bound--free dielectric response.

Fig.~\ref{fig:shell_resolv} shows the individual contributions to
$\varepsilon(q,0)-1$ arising from each occupied shell. Two features are
immediately apparent. First, the largest contributions at low momentum
transfer arise from the valence shells, particularly the $3p$ and $3s$
orbitals: these states are the least tightly bound
and therefore exhibit the largest overlap with continuum final states,
while deep core states are localized near the nucleus and possess larger
excitation energies. Second, all channels eventually decrease with increasing
momentum transfer. This behavior follows directly from
Eq.~(\ref{eq:master}): as $q$ increases, the spherical Bessel
functions oscillate more rapidly and the radial overlap integrals are
suppressed. We find that the large-$q$ asymptotic behavior closely matches the analytic expectation from~Eq.~(\ref{eq:static_dielectric_asymptotic}), now validated numerically in a concrete model. The sum of the core 1$s$, $2$s, and 2$p$ shells gives the semi-analytic curve in Fig.~\ref{fig:epsilon_m1} of the main text.

\end{document}